\documentstyle[12pt]{article}
\def\0{\over } \def\1{\vec }     \def\2{{1\over2}} \def\4{{1\over4}}
\def\5{\bar }  \def\6{\partial } \def\7#1{{#1}\llap{/}}
\def\8#1{{\textstyle{#1}}}       \def\9#1{{\bf {#1}}}
 \def\llp{\hbox to 0pt{\hss /\hskip1.5pt}}
\def\llo{\hbox to 0.2pt{\hss /}} \def\llq{\hbox to 0pt{\hss /\hskip0.5pt}}
\def\so{\supset\hbox to 0pt{\hss $\displaystyle -$\hskip1pt}}

\def\<{\langle } \def\>{\rangle }

\def\bea{\begin{eqnarray}} \def\eea{\end{eqnarray}}
\def\beann{\begin{eqnarray*}} \def\eeann{\end{eqnarray*}}
\def\beq{\begin{equation}} \def\eeq{\end{equation}}

\date{}
\title{
{\large\rm DESY 95-065}\hfill{\large\tt ISSN 0418-9833}\\
{\large\rm April 1995}\hfill\vspace*{3.5cm}\\
Scaling and Diffraction\\
in Deep Inelastic Scattering}
\author{W. Buchm\"{u}ller\\
{\normalsize\it Deutsches Elektronen-Synchrotron DESY, 22603 Hamburg, Germany}
\vspace*{3.5cm}\\
}
\begin{document}

\setlength{\baselineskip}{18pt}
\maketitle
\begin{abstract}
\noindent
We pursue the hypothesis that the events with a large rapidity gap,
observed at HERA, reflect the scattering of electrons off lumps of wee
partons inside the proton. A simple scaling behaviour is predicted
for the diffractive structure functions, which are related to the
inclusive structure function $F_2(x,Q^2)$ at small values of the
scaling variable $x$. The results are compared with
recent measurements of the diffractive structure function
$F_2^D(x,Q^2,M^2)$.
\end{abstract}
\thispagestyle{empty}
\newpage
In the so-called ``rapidity gap'' events, observed and studied at HERA
\cite{zeus}-\cite{h12}, a system of hadrons is observed with
small invariant mass and with a gap in rapidity in the hadronic energy
flow adjacent to the proton beam direction.
This suggests, that in the scattering process a colour neutral part
of the proton with small momentum fraction
is stripped off, which then fragments into the hadrons
visible in the detector. The proton remnant,
carrying most of the energy, escapes undetected close to the
proton direction. The rapidity gap reflects the absense of a colour flow
between proton and current fragments. In analogy to hadronic processes
of similar kind the ``rapidity gap'' events are also referred to as
``diffractive'' events.

It is a remarkable feature of this new class of events, that the cross
section at large momentum transfer $Q^2$ is not suppressed relative to the
total inclusive cross section. Naively, one might expect that the rate for
extracting more than one parton from the proton should rapidly decrease
with increasing $Q^2$. This, however, is not the case.
It is a theoretical challenge to derive the
observed ``leading twist'' behaviour of the diffractive cross section
from QCD, the theory of strong interactions.

In a recent paper \cite{bu}, we have proposed to describe the multi-parton
processes underlying the diffractive events by means of an effective
lagrangian which specifies the coupling between the virtual photon,
the colour singlet wee parton cluster inside the proton
and the hadronic final state. Together with further information on
the cluster density and the mass spectrum of the final
states, one then obtains a prediction for the diffractive
differential cross section. Recently, the H1 collaboration at HERA has
published a first measurement of the diffractive structure function $F_2^D$
for a large range of the kinematic variables \cite{h12}. In this letter,
we therefore extend our previous work \cite{bu} and compare the results with
the recent measurements as well as predictions of
other theoretical approaches.

We consider the inelastic scattering process
\beq \label{process}
e(k)+p(P) \rightarrow e(k')+\tilde{p}(P')+X(P_X)\ ,
\eeq
where $\tilde{p}$ and $X$ denote the proton remnant and the detected
hadronic system, respectively. From the four momenta $k$, $P$,
$P'=(1-\xi)P$, $q=k-k'$ and $P_X=q+\xi P$ one obtains the Lorentz invariant
kinematic variables
\beq
s=(k+P)^2\ ,\ Q^2=-q^2\ ,\  x={Q^2\over 2 q\cdot P}\ ,\
M^2 = (q+\xi P)^2\ .
\eeq
In addition to the first three variables, which characterize ordinary
deep inelastic scattering, the invariant mass $M$ of the detected
hadronic final state occurs as fourth variable. Diffractive events
have been observed for small values of Bjorken's scaling variable
$x$. In this case one has,
\beq \label{invmass}
\xi ={M^2 + Q^2 \over W^2 + Q^2} \simeq x {M^2 + Q^2 \over Q^2}\ ,
\quad W^2 = (q+P)^2 \simeq {Q^2\over x}\ ,
\eeq
where $W$ is the invariant mass of the total hadronic final state
including the proton remnant.

The effective lagrangian used in \cite{bu} is based on the hypothesis
that the wee parton clusters inside the proton can be described by a
scalar field $\sigma$, carrying vacuum quantum numbers. The
hadronic system in the final state is represented by a spectrum
of massive vector states, analogous to generalized vector meson dominance
\cite{yennie}. The coupling between virtual photon, scalar and vector
fields is then given the unique dimension five operator,
\beq \label{effint}
\cal{L}_I = - {1\over 4\Lambda} \sigma(x) F_{\mu\nu}(x) V^{\mu\nu}(x)\ ,
\eeq
where $F_{\mu\nu}=\partial_{\mu}A_{\nu}-\partial_{\nu}A_{\mu}$
and $V_{\mu\nu}=\partial_{\mu}V_{\nu}-\partial_{\nu}V_{\mu}$
are the field strength tensors of the photon field
and the hadronic vector field, respectively. The physical
picture, that the virtual photon acts like a disk of radius $1/Q$
\cite{bj1}, determines the length in eq. (\ref{effint}) as
$1/\Lambda = e\kappa/Q$, where $e$ is the electric charge and $\kappa$
is an unknown constant.

The total diffractive cross section can now be expressed as
\beq\label{xd}
\sigma_D(e p \rightarrow e \tilde{p} X)
= \int dx dQ^2 dM^2 d\xi
{d\hat{\sigma}(e \sigma \rightarrow e V(M))\over d\xi dQ^2}
\rho(Q^2,M^2) f_{\sigma}(\xi,Q^2)\ ,
\eeq
where $d\hat{\sigma}$ is the quasi-elastic cross section for the production
of a vector state of mass $M$, calculated according to eq. (\ref{effint}),
$\rho(Q^2,M^2)$ is the spectral density of vector states,
and $f_{\sigma}(\xi,Q^2)$ is the probability density
of finding a wee parton cluster with momentum fraction
$\xi$ inside the proton. Here we have assumed that the spectral density
only depends on the transverse size and the mass of the produced final
state.

The spectral densities $\rho_T$ and $\rho_L$, assumed in \cite{bu}
for transversely and longitudinally polarized vector states,
were obtained from a fit to the inclusive structure function $F_2$
at small values of $Q^2$ and $M^2$
\cite{sakurai}. In the kinematic range probed at HERA this choice
appears no longer appropriate. In the following, we shall instead use the
ansatz,
\beq\label{density}
\rho_T = \rho_L \equiv \rho = {C\over Q^2+ M^2}\ .
\eeq
This is a simple interpolation between $\rho \propto 1/Q^2$ for $M^2 \ll Q^2$,
and $\rho \propto 1/M^2$ for $Q^2 \ll M^2$, which one may guess based
on dimensional analysis. Note, that at large $M^2$
$\rho$ has to fall off at least as $1/M^2$ in
order to satisfy the unitarity bound
$\sigma_D \leq \sigma_{max} \propto \ln(W^2/Q^2)/Q^2$.
The most general spectral density is obtained by multiplying the
ansatz (\ref{density}) with an arbitrary function of $\beta$, where
\beq
\beta = {Q^2\over Q^2 + M^2}\ .
\eeq
Some examples and their interpretation will be discussed below.

The diffractive cross section is now easily evaluated. For kinematical
reasons, one has
\beq\label{delta}
{d\hat{\sigma}\over d\xi dQ^2}\ \propto\ \delta(\xi - x{Q^2 + M^2\over Q^2})\ .
\eeq
{}From eqs. (\ref{effint}) and (\ref{xd}) one obtains for the
differential cross section in $x$, $Q^2$ and $M^2$ \cite{bu},
\beq
{d\sigma^D \over dx dQ^2 dM^2} = {\pi \alpha^2\kappa^2 \over 4 x Q^4}
\left(1 - y + {y^2\over 2} - 2{Q^2 M^2\over (Q^2+M^2)^2} y^2 \right)
\rho(Q^2,M^2)\xi f_{\sigma}(\xi,Q^2)\ .
\eeq
$\xi$ is now
the function of $x$, $Q^2$ and $M^2$, for which the argument of the
$\delta$-function in eq. (\ref{delta}) vanishes. It corresponds to the
momentum fraction of the parton cluster needed to produce the invariant
mass $M$ (cf. (\ref{invmass})).

Defining transverse and longitudinal diffractive structure functions
in the usual way,
\beq
{d\sigma_D \over dx dQ^2 d\xi} = {4\pi \alpha^2 \over x Q^4}
\left(\left(1 - y + {y^2\over 2}\right)F_2^D(x,Q^2,M^2)
- {y^2\over 2} F_L^D(x,Q^2,M^2)\right)\ ,
\eeq
one obtains ($d\xi = x/Q^2 dM^2$),
\bea
F_2^D(x,Q^2,M^2) &=& {\kappa^2 C \over 16} f_{\sigma}(\xi,Q^2)\ ,\label{fd2}\\
F_L^D(x,Q^2,M^2) &=& 4\beta(1-\beta)F_2^D(x,Q^2,M^2)\ .\label{fdl}
\eea
This result is very simple. Up to an unknown constant, $F_2^D$ is identical
with the probability density for finding a  wee parton cluster with
momentum fraction $\xi$ inside the proton. It is independent
of $\beta$, which is a consequence of our ansatz for the spectral density.
The relation between $F_2^D$ and $F_L^D$ follows from the Lorentz structure
of the effective lagrangian (\ref{effint}) and the assumption
$\rho_T = \rho_L$. The obtained diffractive cross section scales with
$Q^2$, i.e., $\sigma_D \propto \int d\xi F_2^D/Q^2 \propto 1/Q^2$.
Note, that our calculation does not require the proton remnant $\tilde{p}$
to be a proton. Based on ideas of generalized vector meson dominance,
the cross section $\sigma_D(e p \rightarrow e p X)$ has been predicted
to fall off faster than $1/Q^2$ \cite{leo}.

In order to proceed further, we have to determine the probability density
$f_{\sigma}(\xi,Q^2)$. Within the parton model, it appears natural to
build up $f_{\sigma}$ from products of the gluon density $g(x,Q^2)$ and
the sea-quark density $S(x,Q^2)$, which reflect the possible colour singlet
states,
\beq \label{lump}
f_{\sigma}(\xi,Q^2) = \int_{\delta}^{\xi-\delta}d\xi'\left(f_g
g(\xi',Q^2)g(\xi-\xi',Q^2) + f_S S(\xi',Q^2) S(\xi-\xi',Q^2)
+ \ldots\right)\ .
\eeq
Here $f_g$, $f_S$ and $\delta \ll 1$ are constants, and products with
three and more parton densities are represented by the dots.
The parton densities extracted from deep inelastic scattering are singular
at small values of $x$, such that the integral (\ref{lump})
diverges as the infrared cutoff $\delta$ approaches zero. Since the
integral is dominated by the contributions from the two regions
$\xi' \simeq 0$ and $\xi' \simeq \xi$, one obtaines approximately,
\beq\label{lgs}
f_{\sigma}(\xi,Q^2) \simeq \bar{f}_g g(\xi,Q^2) + \bar{f}_S S(\xi,Q^2)
+ \ldots\ ,
\eeq
where the constants $\bar{f}_g$ and $\bar{f}_S$ depend on the cutoff
$\delta$\footnote{In \cite{bu}, the integral (\ref{lump}) was approximated in a
different way which, however, ignored the fact that the dominant contributions
come from the region close to the end points of integration.}.

At small values of $x$, several sets of parton densities \cite{mrs,grv}
show the universal behaviour,
\beq\label{partons}
S(x,Q^2)\ \propto\ g(x,Q^2)\ .
\eeq
For the GRV parton densities, for instance, the ratio $g/S$ is constant
within 10\% for values of $x$ between $10^{-4}$ and $10^{-3}$ \cite{vogt}.
In this case one obtains a simple relation between the diffractive
structure function $F_2^D$ and the inclusive structure function $F_2(x,Q^2)$,
which is proportional to $x S(x,Q^2)$. From eqs. (\ref{fd2}), (\ref{lgs})
and (\ref{partons}) one then obtains the simple scaling relation,
\beq\label{scaling}
F_2^D(x,Q^2,M^2) \simeq {D\over \xi} F_2(\xi,Q^2)\ ,
\eeq
with an unknown constant $D$.
Although the diffractive structure function is defined as a function of
three variables, it only depends on two kinematic variables, the momentum
fraction $\xi$ of the wee parton cluster and $Q^2$. The simple connection
with the inclusive structure function reflects the fact that the momentum
of the parton cluster is essentially carried by a single parton (cf.
(\ref{lgs})). The
remaining ones only screen the colour. The additional factor $1/\xi$
occurs for kinematical reasons.

The scaling relation (\ref{scaling}) can be directly compared with
recent measurements of the two structure functions by the H1 collaboration.
The experimental data for the
diffractive structure function $F_2^D$ are consistent with
\beq
F_2^D(x,Q^2,M^2)\ \propto\  \ln(Q^2)\ \xi^{-n}\ ,
\eeq
where $n=1.19 \pm 0.06 \pm 0.07$ \cite{h12}.
On the other hand, at small values of $x$, the data for
the inclusive structure function can be parametrized as
$F_2(x,Q^2) \propto \ln(Q^2) x^{-d}$, with $d=0.19$ \cite{h13}. These
measurements of $F_2$ and $F_2^D$ are in agreement with the
scaling relation (\ref{scaling}).

The model described above is similar in spirit to
``aligned jet'' models (AJM) \cite{bj2}-\cite{niza}, where the
current fragment is produced by a quark-antiquark pair of transverse
size $1/Q$ penetrating the proton. The predictions of the two approaches
may be compared in terms of the diffractive structure function or,
equivalently, in terms of the diffractive cross section,
conventionally defined as ($x\ll 1$),
\beq
{d\sigma_D\over d M^2}\ =\ {4\pi^2\alpha\over Q^2}
F_2^D(x,Q^2,M^2) {d\xi\over d M^2}\ .
\eeq
{}From eqs. (\ref{fd2}),(\ref{lgs}), and using $g(\xi,Q^2) \sim S(\xi,Q^2) \sim
\xi^{-1-\lambda}$, one obtaines
\beq\label{dsl}
{d\sigma_D\over d M^2}\ \propto\ {1\over Q^2(Q^2 + M^2)} \xi^{-\lambda}\ .
\eeq
In its simplest form, neglecting perturbative QCD corrections \cite{fs},
the AJM predicts the diffractive cross section \cite{bj2},
\beq\label{ajm}
{d\sigma_D^{AJM}\over d M^2}\ \propto\ {1\over (Q^2 + M^2)^2}\ .
\eeq
Integrated over $M^2$, the model yields a total diffractive cross section
$\sigma_D\ \propto\ 1/Q^2$. Note, that this prediction of ``hard diffraction''
as a ``leading twist'' effect was made already before the development of QCD!
The structure function corresponding to the AJM cross section
is $F_2^D \propto \beta/\xi$. In the cluster model described above, the
AJM cross section could be obtained with ``flat'' parton densities
($\lambda=0$) and with the choice of the spectral density
$\rho^{AJM} \propto Q^2/(Q^2 + M^2)^2$ in eq. (\ref{density}).
Another choice, $\rho^{NZ} \propto Q^2 M^2/(Q^2 + M^2)^3$,
would lead to the model of Nikolaev and Zakharov,
which predicts for the diffractive cross section \cite{niza},
\beq\label{nz}
{d\sigma_D^{NZ}\over d M^2}\ \propto\ {M^2\over (Q^2 + M^2)^3}\ ,
\eeq
corresponding to the structure function $F_2^D \propto \beta(1-\beta)/\xi$.

The occurence of jets with large transverse momentum
in diffractive events was first predicted based
on the idea of a ``pomeron structure function'' \cite{ing1}, which can also
account for the HERA results on the diffractive structure function
in deep inelastic scattering
\cite{ing1}-\cite{gestir}. Here, a pomeron structure function
$\tilde{F}_2(\beta,Q^2)$ at some fixed value $Q_0^2$ is needed as
non-perturbative input, where $\beta$ is now interpreted as momentum
fraction of a parton inside the pomeron.
The predictions of the model are then the $Q^2$-dependence of this
pomeron structure function as well as the ``pomeron flux factor'',
which plays the role of the cluster density in the model
described above. More precise measurements of the $\xi$-dependence
and the $\beta$-dependence should be able to distinguish
between the various  models.

Starting from the hypothesis, that the diffractive events in deep inelastic
scattering represent the scattering of electrons off lumps of wee partons,
we have obtained a diffractive structure function which is consistent with
recent measurements. In particular, a simple scaling relation has been
derived between the diffractive and the inclusive structure functions,
which appears to be in agreement with experimental data. This suggests,
that the momentum of the wee parton cluster is essentially carried by
a single parton. Hence, like the inclusive cross section in deep inelastic
electron-proton scattering, also the diffractive cross section may be
essentially due to incoherent electron-parton scattering.
In the case of diffractive processes, however, this interpretation
requires some non-perturbative mechanism of colour screening.

I would like to thank A. Hebecker, G. Ingelman and A. Vogt for valuable
discussions.


\begin{thebibliography}{99}
\bibitem{zeus}
ZEUS collaboration, M. Derrick et al., Phys. Lett. B315 (1993) 481;
B332 (1994) 228
\bibitem{h11}
H1 collaboration, T. Ahmed et al., Nucl. Phys. B429 (1994) 477
\bibitem{h12}
H1 collaboration, T. Ahmed at al., preprint DESY 95-036 (1995)
\bibitem{bu}
W. Buchm\"uller, Phys. Lett. B335 (1994) 479
\bibitem{yennie}
T. H. Baur, R. D. Spital, D. R. Yennie and F. M. Pipkin, Rev. Mod. Phys.
50 (1978) 261
\bibitem{bj1}
J. D. Bjorken, in Proc. of the 1971 Int. Symp. on Electron and Photon
Interactions at High Energies (Ithaca, N. Y., 1971) ed. N. B. Mistry,
p. 13
\bibitem{sakurai}
J. J. Sakurai and D. Schildknecht, Phys. Lett. 42B (1972) 216
\bibitem{leo}
L. Stodolsky, Phys. Lett. B325 (1994) 505
\bibitem{mrs}
A. D. Martin, R. G. Roberts and W. J. Stirling, Phys. Rev. D47 (1993) 867
\bibitem{grv}
M. Gl\"uck, E. Reya and A. Vogt, preprint DESY 94-206 (1994)
\bibitem{vogt}
A. Vogt, private communication
\bibitem{h13}
H1 collaboration, preprint DESY 95-006 (1995)
\bibitem{bj2}
J. D. Bjorken and J. Kogut, Phys. Rev. D8 (1973) 1341
\bibitem{fs}
L. Frankfurt and M. Strikman, Phys. Rep. 160 (1988) 235;\\
H. Abramowicz, L. Frankfurt and M. Strikman, preprint DESY 95-047 (1995)
\bibitem{niza}
N. N. Nikolaev and B. G. Zakharov, Z. Phys. C53 (1992) 331;\\
M. Genovese, N. N. Nikolaev and B. G. Zakharov, preprint DFTT 42/94 (1994)
\bibitem{ing1}
G. Ingelman and P. Schlein, Phys. Lett. 152B (1985) 256
\bibitem{lands}
A. Donnachie and P. V. Landshoff, Phys. Lett. 191B (1987) 309;
Nucl. Phys. B303 (1988) 634
\bibitem{ing2}
G. Ingelman and K. Prytz, Z. Phys. C58 (1993) 285
\bibitem{kohrs}
B. A. Kniehl, H.-G. Kohrs and G. Kramer, Z. Phys. C65 (1995) 657
\bibitem{gestir}
T. Gehrmann and W. J. Stirling, preprint DTP/95/26 (1995)
\end{thebibliography}
\end{document}